\documentclass[twocolumn,english,aps,pra,superscriptaddress,showpacs,tightenlines]{revtex4-1}
\usepackage{amsmath}
\usepackage{graphicx}
\usepackage{amssymb}
\usepackage{CJK}
\usepackage{color}

\begin{document}

\begin{CJK*}{GBK}{song}

\title{Cavity Modified Oscillating Bound States with a $\Lambda$-type giant emitter in a linear waveguide}
\author{Ge \surname{Sun} }
\affiliation{Key Laboratory of Low-Dimension Quantum Structures and Quantum Control of Ministry of Education, Key Laboratory for Matter Microstructure and Function of Hunan Province, Synergetic Innovation Center for Quantum Effects and Applications, Xiangjiang-Laboratory and Department of Physics, Hunan Normal University, Changsha 410081, China}
\affiliation{Institute of Interdisciplinary Studies, Hunan Normal University, Changsha, 410081, China}
\author{Ya \surname{Yang} }
\affiliation{School of Physics and Chemistry, Hunan First Normal University, Changsha 410205, China}
\author{Jing \surname{Li} }
\affiliation{Key Laboratory of Low-Dimension Quantum Structures and Quantum Control of Ministry of Education, Key Laboratory for Matter Microstructure and Function of Hunan Province, Synergetic Innovation Center for Quantum Effects and Applications, Xiangjiang-Laboratory and Department of Physics, Hunan Normal University, Changsha 410081, China}
\affiliation{Institute of Interdisciplinary Studies, Hunan Normal University, Changsha, 410081, China}
\author{Jing \surname{Lu} }
\affiliation{Key Laboratory of Low-Dimension Quantum Structures and Quantum Control of Ministry of Education, Key Laboratory for Matter Microstructure and Function of Hunan Province, Synergetic Innovation Center for Quantum Effects and Applications, Xiangjiang-Laboratory and Department of Physics, Hunan Normal University, Changsha 410081, China}
\affiliation{Institute of Interdisciplinary Studies, Hunan Normal University, Changsha, 410081, China}
\author{Lan \surname{Zhou} }
\thanks{Corresponding author}
\email{zhoulan@hunnu.edu.cn}
\affiliation{Key Laboratory of Low-Dimension Quantum Structures and Quantum Control of Ministry of Education, Key Laboratory for Matter Microstructure and Function of Hunan Province, Synergetic Innovation Center for Quantum Effects and Applications, Xiangjiang-Laboratory and Department of Physics, Hunan Normal University, Changsha 410081, China}
\affiliation{Institute of Interdisciplinary Studies, Hunan Normal University, Changsha, 410081, China}

\begin{abstract}
We study a system composed by a three-level giant atom (3GA), a waveguide initially in the vacuum state, and a single-mode cavity. The 3GA-cavity system is in a strong-coupling regime, and the distance between the coupling points is comparable to the coherent length of a spontaneously emitted photon. The dynamics of the 3GA and its radiative field in the waveguide for long time are analyzed. Besides the steady value, we also found an oscillatory character of the excited state population, a signature of oscillating bound states which is generated by the superposition of two bound states in the continuum. The radiative field propagates in the cavity-like geometry formed by the coupling points. When one bound state is emergent, a sine-like interference pattern is visible for the emitted field intensity in spacetime. An oscillatory character in time and a beat in space for the emitted field intensity are observed when two bound states are emergent in a subspace. The wavelengths and the periods are controlled by the number of the photons in the cavity.
\end{abstract}
\maketitle

\end{CJK*}\narrowtext

\section{Introduction}

Control of the photon-atom interactions lies at the heart of quantum
technologies based on atomic and photonic systems. For example, cavity
quantum electrodynamics (QED) arises due to strong light-matter interactions
by reducing the mode volume of the the electromagnetic (EM) field, and the
strong transverse confinement on the EM fields enhances the light-matter
coupling\cite{RMP89(17),PR718(17)}, then waveguide QED emerges. In
cavity-QED system, quantum emiters (QEs) are usually coupled with a single
cavity mode in a bounded spaces, energies is exchanged periodically between
the QE and the field mode. In contrast, QEs interact with a continuum of
bosonic modes in waveguide-QED system. One of the paradigmatic models is a
QE coupled to a one-dimensional (1D) photonic waveguide, where the QE
undergoes an irreversible exponential decay and a localized photon
wavepacket is generated. The irreversible decay --- spontaneous emission of
the QE's excited state has been exploited to control the single-photon
transport in quantum channels made of one-dimensional (1D) waveguides\cite%
{ShenPRL95,LanPRL101,HoiPRL107,LanPRA89,LanPRA85}. Bound states outside of a
band lead to population trapping within the QE and photon localization
around the QE's position\cite%
{lanPRL111,LuPRA89,SEPRA89,SEPRA96,PRX12,luOL49,LiNJP26}. QEs interacting
via the common vacuum radiation field and a QE in front of a mirror gives
rise to the bound state in the continuum (BIC) due to the appearance of
spontaneous-emission interference\cite%
{LanPRA78,gongPRA78,PRL113,PRA94,PRL124,PRL127,PRA107,DongPRA79,TTCKPRA87,SongCTP69,PRA105Yi}%
. An implicit assumption in conventional light-matter interaction is that
dipoles of QEs are pointlike because atomic dimensions are several orders of
magnitude smaller than the wavelengths of the bosonic modes of the waveguide
they interact with.

Recent and ongoing experiments\cite{Scien346,natcom08,natphys15} on coupling
superconducting artificial atoms to surface acoustic waves inspire the
studies on giant atoms (GAs) whose sizes are comparable or even larger than the
wavelength of the field. One paradigmatic model is the QE coupling to the
waveguide at several points\cite%
{KockPRA90,guoPRA95,wangPRA101,DuPRR03,ChengPRA106,CaiPRA104,GuPRA108}. The
self-interference effect induced by the multi-point coupling leads to some
striking giant-atom effects, such as a frequency-dependent relaxation rate%
\cite{KockPRA90}, waveguide-mediated decoherence-free subspaces\cite%
{NoriPRL120}, tunable chiral bound state\cite{NoriPRL126}. Bound states in
the continuum (BICs) gained significant interest since they provide new
mechanisms to confine radiation, which is of crucial importance for both
fundamental and technological applications. BICs are isolated eigenvalues
embedded in the continuum spectrum of propagating modes of a surrounding space.
In a waveguide-QED system, a BIC is a dressed atom-photon bound state where a
photon is localized in a finite regime and the population is trapped. The
superposition of two BICs generate phenomenon called oscillating bound states%
\cite{guoprr2,KianPRA107,PRA109}, a novel feature of GAs in waveguide QED.
In the paper, we study the emission of a three-level giant atom (3GA) into
a waveguide initially in the vacuum state at two connect points when the
distance between the coupling points is comparable to the coherent length
of a spontaneously emitted photon. The waveguide has a linear dispersion
relation. Two kinds of population trapping owing to the presence of the
dressed atom-photon bound states manifest in the form of a steady value and
a residual oscillatory behavior of the excited state population at long times.
Photon localization occurs in the regime between the two coupling points of a
3GA, and the persistent oscillation of energy in the waveguide trapped between
the two coupling points is generated by the emergence of two dressed atom-photon
bound states under certain condition. We note that while oscillating BICs
have been reported in a waveguide-QED system\cite{guoprr2,KianPRA107,PRA109},
but three coupling points are required at least for a GA coupled to a waveguide.
Our work requires only two coupling points of a GA. In addition, it is possible
for more than two BIC to occur, and the period of the persistent oscillation
at long time is adjustable by the number of the photons in a cavity which is
coupled to one of the transition of the 3GA.

The paper is organized as follows. In Sec.~\ref{Sec:2}, we present the
physical model for the system of a giant three-level quantum emitter (3LE)
with one transition coupled to a 1D semi-infinite waveguide and the other
one coupled to a cavity mode. In Sec.~\ref{Sec:3}, the long-time dynamics of
the 3GA is studied in different subspace by Laplace transformation of the
the delay-differential equation for the 3GA's amplitudes. The appearance of
BICs is understood in terms of interference, and the condition for multiple
BICs are derived in our system. In Sec.~\ref{Sec:4}, photon localization is
studied in terms of the long-time dynamics of the field intensity emitted by
the 3GA. Finally, in Sec.~\ref{Sec:5} we draw our conclusions.


\section{\label{Sec:2}Model And Its Equation of Motion}

\begin{figure}[tbph]
\includegraphics[width=8 cm,clip]{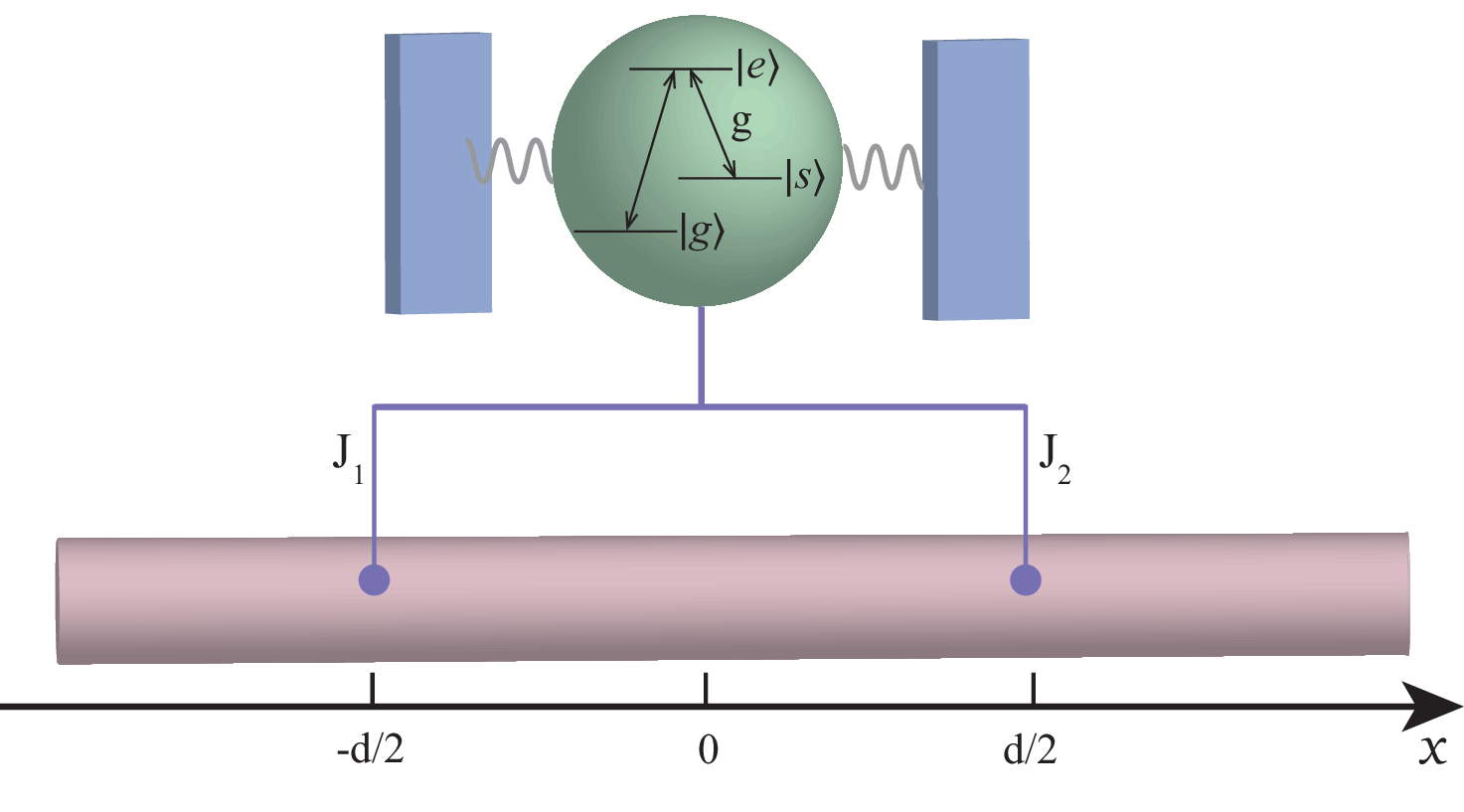}
\caption{Sketch of a three-level GA with one transition $|e\rangle\leftrightarrow |g\rangle$
coupled to a one-dimensional waveguide at $x=\pm d/2$ and the other transition $|e\rangle\leftrightarrow |s\rangle$
interacted with a single-mode cavity.}
\label{fig1}
\end{figure}
The system we studied is shown in Fig.~\ref{fig1}, a 3GA interacts to a 1D
infinite waveguide along the $x$ axis, its relevant transitions are arranged
in a $\Lambda $ configuration, composed of states $|g\rangle $, $|s\rangle $
and $|e\rangle $ with associated state energies $\omega _{g},\omega
_{s}$, and $\omega _{e}$. The 3GA in its excited state $|e\rangle $ can either
transit to a metastable state $|s\rangle $ by a single-mode cavity with a frequency
$\omega _{c}$ via the Jaynes-Cummings (JC) model, or it is transferred to its
ground state $|g\rangle $ at two positions $x=\pm d/2$, from where photons in
the waveguide mode with wave number $k$ are created by the bosonic field operator
$\hat{a}_{k}^{\dagger }$. The full Hamiltonian of the system reads
\begin{eqnarray}
\hat{H} &=&\hat{H}_{0}+g\hat{b}^{\dagger }\left\vert s\right\rangle
\left\langle e\right\vert +h.c.  \label{Eq2-01} \\
&&+\sum_{j=1}^{2}\int dkJ_{j}e^{\left( -1\right) ^{j}\mathrm{i}k\frac{d}{2}}%
\hat{a}_{k}^{\dagger }\left\vert g\right\rangle \left\langle e\right\vert
+h.c.  \notag
\end{eqnarray}%
with the free Hamiltonian
\begin{eqnarray}
\hat{H}_{0} &=&\omega _{s}\left\vert s\right\rangle \left\langle
s\right\vert +\omega _{e}\left\vert e\right\rangle \left\langle e\right\vert
\label{Eq2-02} \\
&&+\omega _{c}\hat{b}^{\dagger }\hat{b}+\int dk\omega _{k}\hat{a}%
_{k}^{\dagger }\hat{a}_{k},  \notag
\end{eqnarray}%
where we select $\left\vert g\right\rangle $ as the reference. The real
coefficient $g$ and $J_{j}=\left\vert J_{j}\right\vert e^{i\varphi _{j}}$
define the coupling strengths of the 3GA to the cavity and the waveguide
with linear dispersion relation $\omega _{k}=v\left\vert k\right\vert $
respectively. No direct coupling is assumed between the cavity and waveguide.
Because the total number of excitations in this system cannot change, the
state at arbitrary moment reads
\begin{eqnarray}
\left\vert \Psi \left( t\right) \right\rangle  &=&\sum_{n}e^{-\mathrm{i}%
n\omega _{c}t}\left[ u_{ne}\left( t\right) \left\vert 0,n,e\right\rangle
+u_{ns}\left( t\right) \left\vert 0,n+1,s\right\rangle \right]   \notag \\
&&+\sum_{n=0}\int_{-\infty }^{\infty }dke^{-\mathrm{i}n\omega _{c}t}\psi
_{nk}\left( t\right) \hat{a}_{k}^{\dagger }\left\vert 0,n,g\right\rangle
\label{Eq2-03}
\end{eqnarray}%
where the ket $\left\vert 0,n,e\right\rangle $ indicates that all modes of
the waveguide are in a vacuum state, the cavity mode is in the number state $%
n$, and the 3GA is in its excited state. The exchange of excitations between
the 3GA and the cavity mode transfer the state $\left\vert 0,n,e\right\rangle $
to state $\left\vert 0,n+1,s\right\rangle $, i.e., it lowers the excited state
of the 3GA to its metastable state and rise the number of photons in the
cavity from $n$ to $n+1$ but keeps the waveguide in its vacuum state, and
vice verse. The 3GA-waveguide coupling makes the 3GA transit from excited state
to its ground state and a photon is emitted into the waveguide mode but keeps
the number of photons in the cavity unchanged, i.e., changes the state
$\left\vert 0,n,e\right\rangle $ to $\hat{a}_{k}^{\dagger }\left\vert 0,n,g\right\rangle $
where state $\hat{a}_{k}^{\dagger }\left\vert 0,n,g\right\rangle $ indicates
that all of the modes of the waveguide are in a vacuum state apart from
mode k. States $\left\vert 0,n,g\right\rangle $ and $\left\vert 0,0,s\right\rangle $
are not coupled to anything else. The corresponding amplitude $u_{ne}\left( t\right) $,
$u_{ns}\left( t\right) $ and $\psi _{nk}\left( t\right) $ satisfies the
normalization condition $1=\sum_{n}\left[ \left\vert u_{ne}\left( t\right)
\right\vert ^{2}+\left\vert u_{ns}\left( t\right) \right\vert ^{2}+\int
dk\left\vert \psi _{nk}\left( t\right) \right\vert ^{2}\right] $. From the
Schr\"{o}dinger equation, the equation governing the dynamics of the total
system reads
\begin{subequations}
\label{Eq2-04}
\begin{eqnarray}
\mathrm{i}\dot{\psi}_{nk}\left( t\right)  &=&\omega _{k}\psi _{nk}\left(
t\right) +\left( J_{1}e^{-\mathrm{i}k\frac{d}{2}}+J_{2}e^{\mathrm{i}k\frac{d%
}{2}}\right) u_{ne}\left( t\right) , \\
\mathrm{i}\dot{u}_{ne}\left( t\right)  &=&\omega _{e}u_{ne}\left( t\right)
+g_{n}u_{ns}\left( t\right)  \\
&&+\int_{-\infty }^{\infty }dk\left( J_{1}^{\ast }e^{\mathrm{i}k\frac{d}{2}%
}+J_{2}^{\ast }e^{-\mathrm{i}k\frac{d}{2}}\right) \psi _{nk}\left( t\right) ,
\notag \\
\mathrm{i}\dot{u}_{ns}\left( t\right)  &=&\left( \omega _{s}+\omega
_{c}\right) u_{ns}\left( t\right) +g_{n}u_{ne}\left( t\right) ,
\end{eqnarray}%
which is a set of coupled equations for the amplitudes in the subspace with $%
n+1$ excitations. Here $g_{n}=g\sqrt{n+1}$ is the $n$-photon Rabi frequency


\section{\label{Sec:3}Time Evolution And Bound States}


We start with the excitations initially in the 3LE and the cavity mode,
which is described by the vector
\end{subequations}
\begin{equation}
\left\vert \Psi \left( 0\right) \right\rangle =\sum_{n}\left[ u_{ne}\left(
0\right) \left\vert 0,n,e\right\rangle +u_{ns}\left( 0\right) \left\vert
0,n+1,s\right\rangle \right] .  \label{Eq3-01}
\end{equation}%
The probability for the 3GA in its excited state regardless of the field
state reads%
\begin{equation}
P_{e}\left( t\right) \equiv \sum_{n}P_{ne}\left( t\right)
=\sum_{n}\left\vert u_{ne}\left( t\right) \right\vert ^{2}.  \label{Eq3-02}
\end{equation}%
To obtain the equation governing the dynamics of the excited state amplitude
$u_{ne}\left( t\right) $, we eliminate the coefficients $\psi _{n,k}\left(
t\right) $ by integrating Eq.~(\ref{Eq2-04}a) in time and substitute the
resulting expression for $\psi _{nk}\left( t\right) $ into Eqs.(\ref{Eq2-04}%
b) and (\ref{Eq2-04}c) with $\psi _{nk}\left( 0\right) =0$, then we arrive
at two coupled differential equations
\begin{subequations}
\label{Eq3-03}
\begin{eqnarray}
\dot{u}_{ne}\left( t\right)  &=&-\mathrm{i}\omega _{e}u_{ne}\left( t\right) -%
\mathrm{i}g_{n}u_{ns}\left( t\right) -\frac{\Gamma }{2}u_{ne}\left( t\right)
\\
&&-\frac{\gamma }{2}u_{ne}\left( t-\tau \right) \Theta \left( t-\tau \right)
,  \notag \\
\dot{u}_{ns}\left( t\right)  &=&-\mathrm{i}\left( \omega _{s}+\omega
_{c}\right) u_{ns}\left( t\right) -\mathrm{i}g_{n}u_{ne}\left( t\right) ,
\end{eqnarray}%
where $\Theta \left( x\right) $ is the Heaviside step function, the damping
parameter $\Gamma =2\pi \left( J_{1}^{\ast }J_{1}+J_{2}^{\ast }J_{2}\right)
/v$ is simply the sum of the decay rate of the single-3GA at each connection
point and the collective damping parameter $\gamma =4\pi \left\vert
J_{1}J_{2}\right\vert \cos \left( \varphi _{1}-\varphi _{2}\right) /v$
depends on the relative phase of the coupling strength at each connection
point. We note that $\Gamma \geq \gamma $. It is convenient to move to the
interaction representation by means of the time dependent transformations
\end{subequations}
\begin{subequations}
\label{Eq3-04}
\begin{eqnarray}
u_{ne}\left( t\right)  &=&U_{ne}\left( t\right) e^{-\mathrm{i}\omega _{e}t},
\\
u_{ns}\left( t\right)  &=&U_{ns}\left( t\right) e^{-\mathrm{i}\omega _{e}t},
\end{eqnarray}%
two coupled differential equations read
\end{subequations}
\begin{subequations}
\label{Eq3-05}
\begin{eqnarray}
\dot{U}_{ne}\left( t\right)  &=&-\mathrm{i}g_{n}U_{ns}\left( t\right) -\frac{%
\Gamma }{2}U_{ne}\left( t\right)    \\
&&-\frac{\gamma }{2}e^{\mathrm{i}\omega _{e}\tau }U_{ne}\left( t-\tau
\right) \Theta \left( t-\tau \right)   \notag \\
\dot{U}_{ns}\left( t\right)  &=&\mathrm{i}\delta U_{ns}\left( t\right) -%
\mathrm{i}g_{n}U_{ne}\left( t\right)
\end{eqnarray}%
with the 3GA-cavity detuning $\delta =\omega _{e}-\omega _{s}-\omega _{c}$.
There are three distinct processes that contribute to the dynamics of the
excited state in Eq.(\ref{Eq3-05}): (1) the 3GA relaxing to its ground state
and emitting a photon in the waveguide characterized by the decay rate $%
\Gamma /2$; (2) a coherent energy exchange between the states $|e\rangle $
and $|s\rangle $ because of its changing excitation with the cavity mode;
(3) the effect of the retarded radiation on the 3GA due to a delay time $%
\tau =d/\upsilon $ that the light emitted from the 3GA into the waveguide
travels along the distance between two connection points. The time dependent
probability amplitude in Eq.(\ref{Eq3-05}) can be calculated by standard
Laplace 
\end{subequations}
\begin{subequations}
\label{Eq3-06}
\begin{eqnarray}
U_{ne} &=&\frac{1}{2\pi \mathrm{i}}\int \frac{dse^{st}\left[ \left( s-%
\mathrm{i}\delta \right) U_{ne}\left( 0\right) -\mathrm{i}g_{n}^{\ast
}U_{ns}\left( 0\right) \right] }{\left[ s+\frac{\Gamma }{2}+\frac{\gamma }{2}%
e^{\left( \mathrm{i}\omega _{e}-s\right) \tau }\right] \left( s-\mathrm{i}%
\delta \right) +\left\vert g_{n}\right\vert ^{2}} \\
U_{ns} &=&\frac{1}{2\pi \mathrm{i}}\int \frac{dse^{st}\left[ s+\frac{\Gamma
}{2}+\frac{\gamma }{2}e^{\left( \mathrm{i}\omega _{e}-s\right) \tau }\right]
U_{ns}\left( 0\right) }{\left[ s+\frac{\Gamma }{2}+\frac{\gamma }{2}%
e^{\left( \mathrm{i}\omega _{e}-s\right) \tau }\right] \left( s-\mathrm{i}%
\delta \right) +\left\vert g_{n}\right\vert ^{2}} \\
&&-\frac{1}{2\pi \mathrm{i}}\int \frac{dse^{st}\mathrm{i}g_{n}U_{ne}\left(
0\right) }{\left[ s+\frac{\Gamma }{2}+\frac{\gamma }{2}e^{\left( \mathrm{i}%
\omega _{e}-s\right) \tau }\right] \left( s-\mathrm{i}\delta \right)
+\left\vert g_{n}\right\vert ^{2}}  \notag
\end{eqnarray}%
In the case of $g=0$ and $\tau \rightarrow \infty $, the system become a
point-like two-level atom interacting with a waveguide, so the probability
in the metastable state remain its initial value, but the point-like
two-level atom decays exponentially with rate $\Gamma $ to its ground state
accompanied by an irreversible release of energy to the vacuum of a
waveguide, which can be obtained from Eq.(\ref{Eq3-06}) by letting $g=0$ and
$\gamma =0$. In the case of $g=0$, the metastable state involves freely, the
probability amplitude
\end{subequations}
\begin{equation}
U_{ne}\left( t\right) =\int ds\frac{e^{st}U_{ne}\left( 0\right) /\left( 2\pi
\mathrm{i}\right) }{s+\frac{\Gamma }{2}+\frac{\gamma }{2}e^{-s\tau }e^{%
\mathrm{i}2\pi d/\lambda _{e}}},  \label{Eq3-07}
\end{equation}%
which describes the dynamics of a two-level giant atom (2GA) interacting
with a waveguide at two connection points\cite{guoprr2}. In the interval $%
t\in \left( 0,\tau \right) $, the 2GA decays exponentially with rate $\Gamma
$. As time increases, its dynamics is determined by three parameters: the
characteristic wavelength $\lambda _{e}=2\pi v/\omega _{e}$, the distance $%
d=v\tau $ among the two connection points and the coherence length $%
L_{c}=v/\Gamma $. For GA, $\lambda _{e}<d,L_{c}$ is usually satisfied. If
two coupling points are close enough that we can ignore any effects resulting
from spatial distance, i.e., $d\ll L_{c}$, the factor $e^{-s\tau }$ in
Eq.(\ref{Eq3-07}) is set to one. But it is impossible to assume that
factor $e^{\mathrm{i}2\pi d/\lambda _{e}}$ have the same value since the
distances is at least of the order of a resonant wavelength. Therefore, a
bound state is emergent at $d=\left( m+1/2\right) \lambda _{e}$ and
$\Gamma =\gamma $ or $d=m\lambda _{e}$ and $\Gamma =-\gamma $.

\begin{figure}[tbph]
\includegraphics[width=8 cm,clip]{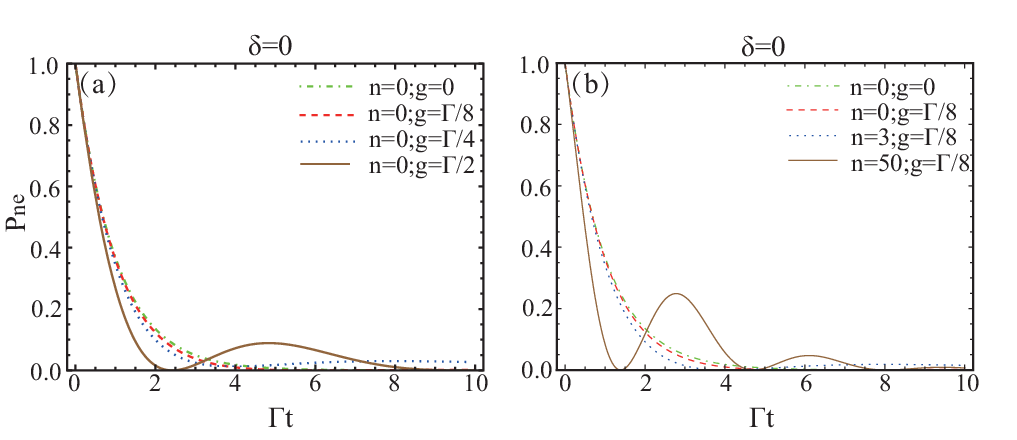}
\caption{(color online)Time evolution of populations $P_{ne}$ as a function
of scaled time $\Gamma t$ for $\protect\delta =0$ with initial condition $%
u_{ne}\left( 0\right) =1,u_{nc}\left( 0\right) =0$ when $\protect\tau %
\rightarrow \infty $ (corresponding to a point-like 3LE).}
\label{fig2}
\end{figure}

When $\tau \rightarrow \infty $, equation~(\ref{Eq3-06}) depicts the
dynamics of a point-like three-level atom (3LA) coupled to a 1D infinite
waveguide. In Fig.~\ref{fig2}, we have plotted the population $P_{ne}\left(
t\right) $ on the excited level as a function of the scaled time $\Gamma t$
with the initial condition $u_{ne}\left( 0\right) =1,u_{ns}\left( 0\right) =0
$ when $\tau \rightarrow \infty $.
Clearly, its dynamics is strongly dependent on the 3LA-cavity coupling strength
and the number of the photons in the cavity mode. In Fig.\ref{fig2}, the green
dotted-dashed line depicts the exponential decay of the point-like 2LA for
comparison. It can be found that the population $P_{ne}$ decreases faster
than that of 2LA in the regime $g_{n}\in \left( 0,\Gamma /4\right) $, the no
coherent evolution indicates that the 3LA-cavity system is in weak coupling
regime. In the regime $g_{n}\geq \Gamma /4$, signatures of the coherent
evolution --- oscillations can be observed, coherent evolution dominates for
some time until dissipation destroys it. The coherent evolution indicates
that the 3LA-cavity system is in strong coupling regime. In such regime,
the eigenstates of the 3LA-cavity are
\begin{subequations}
\label{Eq3-08}
\begin{eqnarray}
\left\vert n_{+}\right\rangle  &=&\frac{g_{n}}{\sqrt{2\Delta _{n}\omega
_{n+}^{\prime }}}\left\vert n,e\right\rangle +\sqrt{\frac{\omega
_{n+}^{\prime }}{2\Delta _{n}}}\left\vert n+1,s\right\rangle  \\
\left\vert n_{-}\right\rangle  &=&\frac{g_{n}}{\sqrt{-2\Delta _{n}\omega
_{n-}^{\prime }}}\left\vert n,e\right\rangle -\sqrt{\frac{-\omega
_{n-}^{\prime }}{2\Delta _{n}}}\left\vert n+1,s\right\rangle
\end{eqnarray}%
with the corresponding energies $\omega _{n\pm }=\omega _{e}+\omega _{n_{\pm
}}^{\prime }$, where $\omega _{n_{\pm }}^{\prime }=-\delta /2\pm \Delta _{n}$
and $\Delta _{n}=\sqrt{g_{n}^{2}+\delta ^{2}/4}$ determines the time scale
of energy exchange. The detuning $\delta $ increases the splitting between
the dressed state in Eq.(\ref{Eq3-08}). For $\delta \gg \left\vert
g\right\vert $, the dressed states $\left\vert n_{\pm }\right\rangle $ reduce to
the product states $\left\vert n,e\right\rangle $ and $\left\vert
n+1,s\right\rangle $. In the following discussion, we will assume that the
3LA-cavity system is in the strong coupling regime at $\delta =0$. We
emphasize that these assumptions do not limit qualitatively the physics of
the system.
\begin{figure*}[tbp]
\includegraphics[width=0.9 \textwidth]{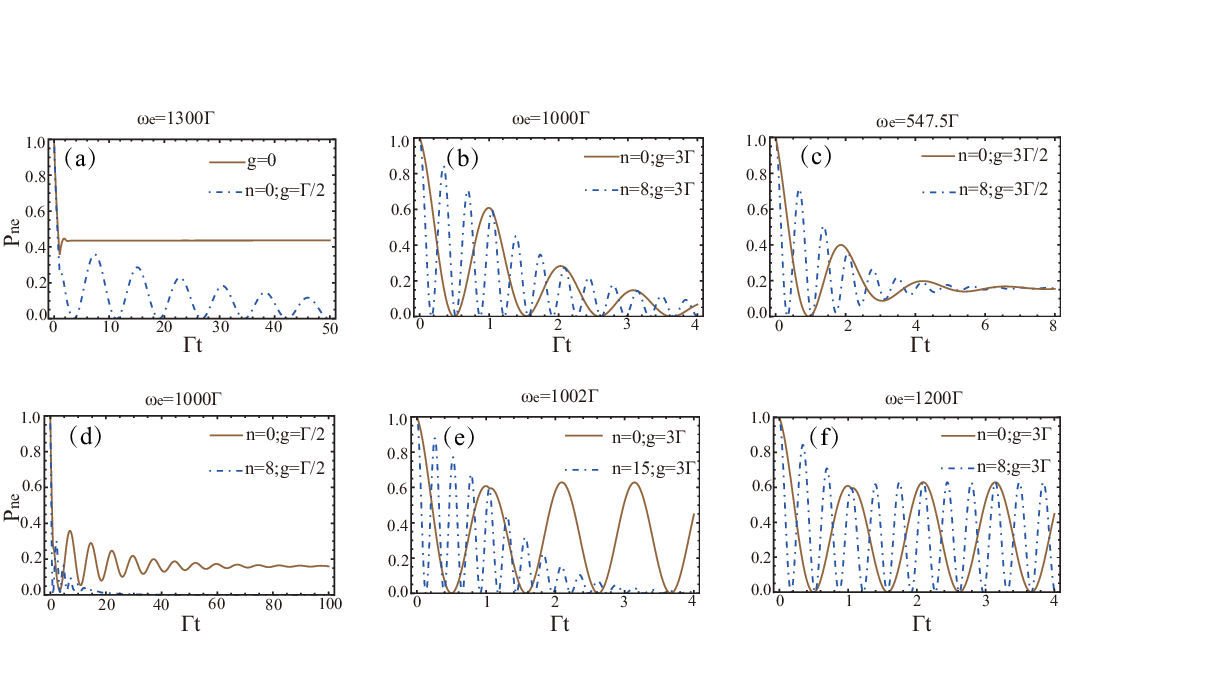}
\caption{(color online)Time evolution of populations $P_{ne}$ for $\protect%
\delta =0$ and $\protect\gamma =\Gamma $ with initial condition $%
u_{ne}\left( 0\right) =1,u_{ns}\left( 0\right) =0$ when $d\sim L_{c}$.}
\label{fig3}
\end{figure*}

When $\tau $ is finite, a photon emitted from the 3GA into the waveguide via
a coupling point will propagate to the left and right,
some part will be absorbed by the 3GA at the other coupling point until time
$\tau $, alternatively some may also be reflected back by the other coupling
point to its original coupling point. Then the photon is emitted again from
the 3GA. The process of absorption and reflection would be repeated at times
$t=n\tau $, which consequently alters the evolution and therefore the rate
of radiation. In Fig.\ref{fig3}, the time evolution of $P_{ne}\left(
t\right) $ is shown by numerically solving Eq.(\ref{Eq3-03}) with the
distance $d$ comparable to the coherent length $L_{c}$ which are all larger
than an optical wavelength $\lambda _{e}$. Here, the dynamics starts in the
state $\left\vert n,e\right\rangle $. The time evolution of the 2GA's
population with $g=0$ is displayed with solid line for comparison in Fig.\ref%
{fig3}a. It is visible from Fig.\ref{fig3}a that the bound state occurring
at $d=(m+1/2)\lambda _{e}$ partially traps the energy in the 2GA, however,
the population of the 3GA's excited state relaxes to its ground state, which
indicates that a bound state is impossible to be formed at $d=(m+1/2)\lambda
_{e}$. A periodic modulation superimposed on exponential decay can be
observed from Fig.\ref{fig3}b in the subspaces with excitation number $N=1$
and $9$ respectively. A considerable trapping of the energy within the
3GA-cavity for long times is clearly visible with a steady value of
population in Fig.\ref{fig3}c-d and the stationary oscillation in Fig.\ref%
{fig3}d-f. To understand the considerable trapping of excitation, let us
find the poles of Eq.(\ref{Eq3-06}). The pure imaginary poles $s=-\mathrm{i}%
\omega $ require
\end{subequations}
\begin{subequations}
\label{Eq3-09}
\begin{eqnarray}
0 &=&\left\{ \frac{\gamma }{2}\sin \left[ \left( \omega +\omega _{e}\right)
\tau \right] -\omega \right\} \left( \omega +\delta \right) +g_{n}^{2} \\
0 &=&-\mathrm{i}\left[ \frac{\Gamma }{2}+\frac{\gamma }{2}\cos \left[ \left(
\omega +\omega _{e}\right) \tau \right] \right] \left( \omega +\delta
\right)
\end{eqnarray}%
to be satisfied. Solving the two equations in Eq.(\ref{Eq3-09}), we obtain $%
\omega =\omega _{n_{\pm }}^{\prime }\ $and $\left( \omega +\omega
_{e}\right) \tau =\left( 2m+1\right) \pi $ ($\left( \omega +\omega
_{e}\right) \tau =2m\pi $) for $\Gamma =\gamma $ ($\Gamma =-\gamma $), i.e.,
It is possible to form a bound state once the distance $d=\left(
m+1/2\right) \lambda _{n+}$ ($\left( m+1/2\right) \lambda _{n-}$) with $%
\lambda _{n\pm }=2\pi v/\omega _{n\pm }$ and $\Gamma =\gamma $ or $%
d=m\lambda _{n+}$ ($m\lambda _{n-}$) and $\Gamma =-\gamma $, as a result,
the population stays in one of the dressed states. Moreover, it is possible
to find two integers $m_{n}^{\pm }$ satisfying $d=\left(
q_{n}^{+}+1/2\right) \lambda _{n+}=\left( q_{n}^{-}+1/2\right) \lambda _{n-}$
($d=q_{n}^{+}\lambda _{n+}=q_{n}^{-}\lambda _{n-}$) simultaneously when $%
\Gamma =\gamma $ ($\Gamma =-\gamma $), i.e., $d=\left(
q_{n}^{+}-q_{n}^{-}\right) \lambda _{nb}$ with $\lambda _{nb}=2\pi v/\left(
\omega _{n+}-\omega _{n-}\right) $, population is trapped in both dressed
states. After all unstable state die out, the amplitude becomes
\end{subequations}
\begin{subequations}
\label{Eq3-10}
\begin{eqnarray}
u_{ne}\left( t\right)  &=&\frac{e^{-\mathrm{i}\omega _{n_{+}}t}2\cos
^{2}\theta _{n}}{\gamma \tau \cos ^{2}\theta _{n}+2}\left[ u_{ne}\left(
0\right) +\frac{2g_{n}u_{ns}\left( 0\right) }{\delta +2\Delta _{n}}\right]
\\
&&+\frac{e^{-\mathrm{i}\omega _{n_{-}}t}2\sin ^{2}\theta _{n}}{\gamma \tau
\sin ^{2}\theta _{n}+2}\left[ u_{ne}\left( 0\right) +\frac{%
2g_{n}u_{ns}\left( 0\right) }{\delta -2\Delta _{n}}\right]
\end{eqnarray}%
where $\sin \theta _{n}=\sqrt{\frac{2\Delta _{n}-\delta }{4\Delta _{n}}}$and
$\cos \theta _{n}=\sqrt{\frac{2\Delta _{n}+\delta }{4\Delta _{n}}}$. So one
observes that the population $P_{ne}$ oscillates with time and the frequency
$2\Delta _{n}$ of the oscillations is equal to the frequency difference $%
\left\vert \omega _{n+}-\omega _{n-}\right\vert $ between two dressed
states. We would like to write $u_{ne}\left( t\right) $ as the sum of $%
u_{ne}^{\pm }\left( t\right) $ with
\end{subequations}
\begin{subequations}
\label{Eq3-11}
\begin{eqnarray}
u_{ne}^{+}\left( t\right)  &=&\frac{e^{-\mathrm{i}\omega _{n_{+}}t}2\cos
^{2}\theta _{n}}{\gamma \tau \cos ^{2}\theta _{n}+2}\left[ u_{ne}\left(
0\right) +\frac{2g_{n}u_{ns}\left( 0\right) }{\delta +2\Delta _{n}}\right] ,
\\
u_{ne}^{-}\left( t\right)  &=&\frac{e^{-\mathrm{i}\omega _{n_{-}}t}2\sin
^{2}\theta _{n}}{\gamma \tau \sin ^{2}\theta _{n}+2}\left[ u_{ne}\left(
0\right) +\frac{2g_{n}u_{ns}\left( 0\right) }{\delta -2\Delta _{n}}\right] .
\end{eqnarray}%
Obviously, $u_{ne}^{\pm }\left( t\right) $ is the amplitude of the dressed
states in Eq.(\ref{Eq3-08}) since it evolves at the frequencies $\omega
_{n_{\pm }}$.

Figure~\ref{fig3}d presents a nonzero steady value of population in $N=1$
subspace and a decay in $N=9$ subspace, however, it is possible for one
bound state to occur in different subspaces, see the steady-state 3GA-cavity
population in both $N=1$ subspace and $N=9$ subspace in Fig.\ref{fig3}c. The
condition for the coexistence of a bound state in different subspace is $%
d=2\pi v\left( q_{m}^{\alpha }-q_{n}^{\beta }\right) /\left( \alpha
g_{m}-\beta g_{n}\right) $ with $\alpha ,\beta \in \left\{ \pm \right\} $,
where $q_{m}^{\alpha }$ and $q_{n}^{\beta }$ are the integers in $m+1$
subspace and $n+1$ subspace respectively. Figure~\ref{fig3}e presents a
stationary oscillatory behavior in $N=1$ subspace and a decay in $N=15$
subspace, however, one may also find the stationary oscillatory behavior in
both $N=1$ subspace and $N=9$ subspace, and oscillations with different
periods in different subspace as shown in Fig.\ref{fig3}f. To keep the
constant level of oscillation in $n+1$ subspace, integers $q_{n}^{\pm }$
must already exist, so do integers $q_{m}^{\pm }$ in $m+1$ subspace. The
physical origins of the bound states formation ask integers $q_{m}^{\pm }$
and $q_{n}^{\pm }$ to satisfy the following relation
\end{subequations}
\begin{equation}
\frac{\Delta _{n}}{\Delta _{m}}=\frac{q_{n}^{+}-q_{n}^{-}}{%
q_{m}^{+}-q_{m}^{-}}
\end{equation}%
in order to emerge constant level of oscillation simultaneously in two
subspaces.

\section{\label{Sec:4}Emitted photonic modes}

\begin{figure*}[tbp]
\includegraphics[width=0.9 \textwidth]{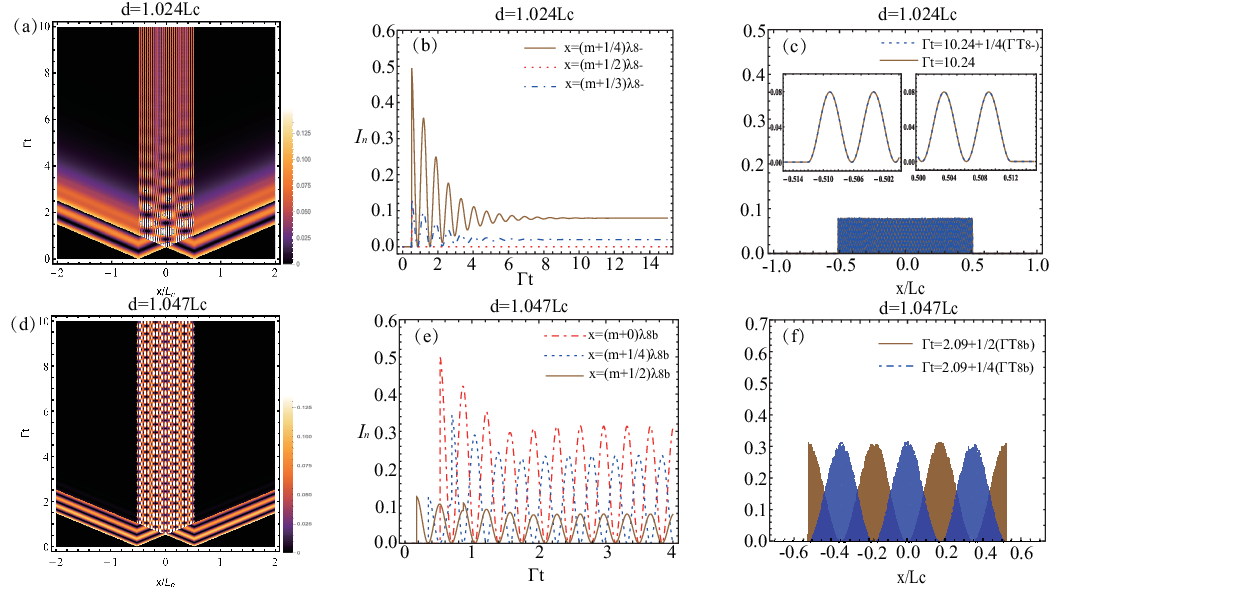}
\caption{(color online)The radiative intensity $I_{n}$ for $\protect\delta =0
$ and $\protect\gamma =\Gamma $ with initial condition $u_{ne}\left(
0\right) =1,u_{ns}\left( 0\right) =0$ when $d\sim L_{c}$.}
\label{fig4}
\end{figure*}

To get a more physical insight in the steady state of the dynamics, it is
instructive to explore the waveguide-QED dynamics from the perspective of
the emitted field intensity. We consider the radiative intensity $I\left(
x,t\right) =\left\langle \Psi \left( t\right) \right\vert \hat{E}^{-}\left(
x\right) \hat{E}^{+}\left( x\right) \left\vert \Psi \left( t\right)
\right\rangle $ detected at a point $x$ at the moment of time $t$, where the
negative frequency parts of the waveguide field in the Schr\"{o}dinger
picture
\begin{equation*}
\hat{E}^{+}\left( x\right) =\frac{1}{\sqrt{2\pi }}\int dke^{ikx}\hat{a}_{k}.
\end{equation*}%
Calculations are limited to the steady-state intensity under the condition $%
J_{1}=J_{2}$ and $\varphi _{1}-\varphi _{2}=n\pi $. Note that there is only
one photon in the waveguide, the intensity can be rewritten as $I\left(
x,t\right) =\sum_{n}I_{n}\left( x,t\right) =\sum_{n}\left\vert \Psi
_{n}\left( x,t\right) \right\vert ^{2}$ with%
\begin{eqnarray*}
\Psi _{n}\left( x,t\right)  &=&-\mathrm{i}e^{\mathrm{i}\varphi _{1}}\sqrt{%
\frac{\Gamma }{8v}}u_{ne}\left( t-\left\vert \tau _{x}^{-}\right\vert
\right) \Theta \left( t-\left\vert \tau _{x}^{-}\right\vert \right)  \\
&&-\mathrm{i}e^{\mathrm{i}\varphi _{2}}\sqrt{\frac{\Gamma }{8v}}u_{ne}\left(
t-\left\vert \tau _{x}^{+}\right\vert \right) \Theta \left( t-\left\vert
\tau _{x}^{+}\right\vert \right)
\end{eqnarray*}%
where $\tau _{x}^{\pm }=x/v\pm \tau /2$. Figure~\ref{fig4} numerically shows
the dependence of the intensity $I_{n}\left( x,t\right) $ on time and
coordinator for $d\sim L_{c}$ at $\varphi _{1}-\varphi _{2}=0$ in $N=9$
subspace. A right-going (left-going) wave propagates far away from the 3GA
at the coupling point $x=d/2$ ($x=-d/2$), and some photonic wave propagates
back and forth between the coupling points while its amplitude is damped due
to energy exchanged between the 3GA and the waveguide. The back and forth
waves are superimposed in the regime between the coupling points, see a
series of alternating bright and dark fringes in Fig.~\ref{fig4}a or bright
and dark regions in Fig.~\ref{fig4}d. Figure~\ref{fig4}b shows that after an
initial rapid changes with time, the intensity detected at a point $x$ is
eventually a time-independent value. The intensity in space for different
times in Fig.~\ref{fig4}c presents an oscillating wave fixed in space,
indicating that the field comes to a time-independent steady state and nodes
are the coupling points. However, an oscillation fixed in time is displayed
in the long time as shown in Fig.~\ref{fig4}e, and the position of the
detector modifies the depth of the intensity. Figure ~\ref{fig4}f shows a
behavior of a periodic modulation superimposed on periodic modulation
indicating superposition of waves with different wavelengths.
The residual field intensity mainly comes from steady states. If the system
have residues in two BICs at frequency $\omega _{n+}$ and $\omega _{n-}$, we
can write $\Psi _{n}\left( x,t\right) =\Psi _{n}^{+}\left( x,t\right) +\Psi
_{n}^{-}\left( x,t\right) $ as a superposition of two amplitudes $\Psi
_{n}^{\pm }\left( x,t\right) $ at position $x$ of two BICs with%
\begin{eqnarray*}
\Psi _{n}^{\pm }\left( x,t\right)  &=&e^{\mathrm{i}\varphi _{1}}\sqrt{\frac{%
\Gamma }{2v}}u_{ne}^{\pm }\left( t\right) \sin \frac{\omega _{n_{\pm
}}\left( d-2x\right) }{2v}\text{, }x\in \left[ -\frac{d}{2},\frac{d}{2}%
\right]  \\
\Psi _{n}^{\pm }\left( x,t\right)  &=&0\text{, otherwise }
\end{eqnarray*}%
This means that the field intensity oscillates sinusoidally with period $%
T_{nb}=2\pi /\left\vert \omega _{n+}-\omega _{n-}\right\vert $ in a given
subspace, so an oscillating BIC is exhibited in our system. For a given
time, two light waves of different wavelengths interfere, which produces
beats in real space. If only one BIC appears, the intensity in the long-time
limit is time independent, and its interference pattern is sine-like in
space.


\section{\label{Sec:5}Conclusion}


We have studied the dynamics of a 3GA in $\Lambda $ configuration with one
transition interacted at two points with a waveguide initially in the vacuum
state, and the other transition coupled to a single-mode cavity. We focus on the
regime where the distance $d$ between the coupling points is comparable to
the coherent length $L_{c}$ of a spontaneously emitted photon and the
coupling of the 3GA and cavity is strong. The presence of the cavity changes
the number of the transition interacted with a waveguide from one in
bare-state basis to at least two in the dressed-state basis. The emergence
of dressed bound states in each subspace is determined by the coupling
strengths between the 3GA and the waveguide and the energies of the dressed
states. The signature of one bound state in a subspace is displayed by a steady
value for the dynamics of the 3GA in the long-time limit, and a sine-like
interference pattern located in the regime between the coupling points for
the 3GA's radiative intensity in spacetime. The coexistence of the two bound
states in a subspace is visible by an oscillatory character in time and a beat
in space for the 3GA's radiative intensity, and an oscillatory character for
the dynamics of the 3GA in the long-time limit. The number of photons in the
cavity modifies the wavelengths and the periods of the field intensity, so
does the periods of oscillation for population of the 3GA's excite state.

\begin{acknowledgments}
This work was supported by NSFC Grants No.11935006, No. 12075082, No. 12247105,
the science and technology innovation Program of Hunan Province (Grant No. 2020RC4047),
XJ2302001, and the Science $\And$ Technology Department of Hunan Provincial Program (2023ZJ1010).
\end{acknowledgments}


\begin{thebibliography}{99}
\bibitem{RMP89(17)} D. Roy, C.M. Wilson, and O. Firstenberg, Colloquium:
Strongly interacting photons in one-dimensional continuum, Rev. Mod. Phys.
89, 021001 (2017).

\bibitem{PR718(17)} X. Gu, A. F. Kockum, A. Miranowicz, Y.-X. Liu, and F.
Nori, Microwave photonics with superconducting quantum circuits, Phys. Rep.
718--719, 1 (2017).

\bibitem{ShenPRL95} J. T. Shen and S. Fan, Coherent Single Photon Transport
in a One-DimensionalWaveguide Coupled with Superconducting Quantum Bits,
Phys. Rev. Lett. 95, 213001 (2005).

\bibitem{LanPRL101} L. Zhou, Z. R. Gong, Y.-X. Liu, C. P. Sun, and F. Nori,
Controllable Scattering of a Single Photon inside a One-Dimensional
Resonator Waveguide,\ Phys. Rev. Lett. 101, 100501 (2008).

\bibitem{HoiPRL107} I.-C. Hoi, C. M. Wilson, G. Johansson, T. Palomaki, B.
Peropadre, and P. Delsing, Demonstration of a Single-Photon Router in the
Microwave Regime, Phys. Rev. Lett. 107, 073601 (2011).

\bibitem{LanPRA89} Z. H. Wang, L. Zhou, Y. Li, and C. P. Sun, Controllable
single-photon frequency converter via a one-dimensional waveguide, Phys.
Rev. A 89, 053813 (2014).

\bibitem{LuOE23} J. Lu, Z. H. Wang, and L. Zhou, T-shaped single-photon
router, Opt. Exp. 23, 22955 (2015).

\bibitem{YaAPL123} Y.-K. Luo, Ya Yang, Jing Lu, and Lan Zhou, Control of a
single-photon router via an extra cavity, Appl. Phys. Lett. 123, 211103
(2023).

\bibitem{YaAPL124} Y. Yang, J. Lu, and Lan Zhou, Controllable nonreciprocal
single-photon frequency converter via a four-level system, Appl. Phys. Lett.
124, 141102 (2024).

\bibitem{LanPRA85} L. Zhou, Y. Chang, H. Dong, L.-M. Kuang, and C. P. Sun,
Inherent Mach-Zehnder interference with "which-way" detection for single
particle scattering in one dimension, Phys. Rev. A 85, 013806 (2012).

\bibitem{lanPRL111} L. Zhou, L. P. Yang, Y. Li, and C. P. Sun, Quantum
Routing of Single Photons with a Cyclic Three-Level System, Phys. Rev. Lett.
111, 103604 (2013).

\bibitem{LuPRA89} J. Lu, L. Zhou, L. M. Kuang, and F. Nori, Single-photon
router: Coherent control of multichannel scattering for single photons with
quantum interferences, Phys. Rev. A 89, 013805 (2014).

\bibitem{SEPRA89} F. Lombardo, F. Ciccarello, and G. M. Palma, Photon
localization versus population trapping in a coupled-cavity array, Phys.
Rev. A 89, 053826 (2014).

\bibitem{SEPRA96} E. S\'{a}chez-Burillo, et al., Dynamical signatures of
bound states in waveguide QED, Phys. Rev. A 96, 023831 (2017).

\bibitem{PRX12} M. Scigliuzzo, G. Calaj, F. Ciccarello et.al, Controlling
Atom-Photon Bound States in an Array of Josephson-Junction Resonators, Phys.
Rev. X 12, 031036 (2022).

\bibitem{luOL49} Z.L. Lu, J. Li,, J. Lu, and L. Zhou, Controlling
atom-photon bound states in a coupled resonator array with a two-level
quantum emitter, Optics Letters, 49, 806 (2024).

\bibitem{LiNJP26} J. Li, J. Lu, Z. R. Gong and L. Zhou, Tunable chiral bound
states in a dimer chain of coupled resonators, New J. Phys. 26, 033025
(2024).

\bibitem{LanPRA78} L. Zhou, H. Dong, Y.-X. Liu, C. P. Sun, and F. Nori,
Quantum supercavity with atomic mirrors, Phys. Rev. A 78, 063827 (2008).

\bibitem{gongPRA78} Z. R. Gong, H. Ian, L. Zhou, and C. P. Sun, Controlling
quasibound states in a one-dimensional continuum through an
electromagnetically-induced-transparency mechanism, Phys. Rev. A 78, 053806
(2008).

\bibitem{PRL113} F. Fratini, E. Mascarenhas, L. Safari, J.-P. Poizat, D.
Valente, A. Auff`eves, D. Gerace, and M. F. Santos, Fabry-Perot
Interferometer with Quantum Mirrors: Nonlinear Light Transport and
Rectification, Phys. Rev. Lett. 113, 243601 (2014).

\bibitem{PRA94} P. Facchi, M. S. Kim, S. Pascazio, F. V. Pepe, D. Pomarico,
and T. Tufarelli, Bound states and entanglement generation in waveguide
quantum electrodynamics, Phys. Rev. A 94, 043839 (2016)

\bibitem{PRL124} K. Sinha, P. Meystre, E.A. Goldschmidt, F.K. Fatemi, S. L.
Rolston, and P. Solano, Non-Markovian Collective Emission from
Macroscopically Separated Emitters, Phys. Rev. Lett. 124, 043603 (2020)

\bibitem{PRL127} H. S. Han, A. Lee, K, Sinha, F, K. Fatemi, and S. L.
Rolston, Observation of Vacuum-Induced Collective Quantum Beats, Phys. Rev.
Lett. 127, 073604 (2021).

\bibitem{PRA107} P. Solano, P. Barberis-Blostein, and K. Sinha, Dissimilar
collective decay and directional emission from two quantum emitters, Phys.
Rev. A 107, 023723 (2023)

\bibitem{DongPRA79} H. Dong, Z. R. Gong, H. Ian, L. Zhou, and C. P. Sun,
Intrinsic cavity QED and emergent quasinormal modes for a single photon,
Phys. Rev. A 79, 063847 (2009).

\bibitem{TTCKPRA87} T. Tufarelli, F. Ciccarello, and M. S. Kim, Dynamics of
spontaneous emission in a single-end photonic waveguide, Phys. Rev. A 87,
013820 (2013).

\bibitem{SongCTP69} H. X. Song, X. Q. Sun, J. Lu, and L. Zhou, Spatial
dependent spontaneous emission of an atom in a semi-infinite waveguide of
rectangular cross section, Commun. Theor. Phys. 69, 59 (2018).

\bibitem{PRA105Yi} L. Xin,, S. Xu, X. X. Yi, and H. Z. Shen, Tunable
non-Markovian dynamics with a three-level atom mediated by the classical
laser in a semi-infinite photonic waveguide, Phys. Rev. A 105, 053706 (2022).

\bibitem{Scien346} M. V. Gustafsson, T. Aref, A. F. Kockum, M. K. Ekstr\"{o}%
m, G. Johansson, and P. Delsing, Propagating phonons coupled to an
artificial atom, Science 346, 207 (2014).

\bibitem{natcom08} R. Manenti, A. F. Kockum, A. Patterson, T. Behrle, J.
Rahamim, G. Tancredi, F. Nori, and P. J. Leek, Circuit quantum
acoustodynamics with surface acoustic waves, Nat. Commun. 8, 975 (2017).

\bibitem{natphys15} G. Andersson, B. Suri, L. Guo, T. Aref, and P. Delsing,
Nonexponential decay of a giant artificial atom, Nat. Phys. 15, 1123 (2019).

\bibitem{KockPRA90} A. F. Kockum, P. Delsing, and G. Johansson, Designing
frequency-dependent relaxation rates and Lamb shifts for a giant artificial
atom, Phys. Rev. A 90, 013837 (2014)

\bibitem{guoPRA95} L.z. Guo, A. Grimsmo, A. F. Kockum, M. Pletyukhov and G.
Johansson, Giant acoustic atom: A single quantum system with a deterministic
time delay, Phys. Rev. A 95, 053821 (2017).

\bibitem{wangPRA101} W. Zhao and Z. Wang, Single-photon scattering and bound
states in an atom-waveguide system with two or multiple coupling points,
Phys. Rev. A 101, 053855 (2020).

\bibitem{DuPRR03} Lei Du, Y.-T. Chen, and Y. Li, Nonreciprocal frequency
conversion with chiral $\Lambda $-type atoms, Phys. Rev. Res. 3, 043226
(2021)

\bibitem{ChengPRA106} W. Cheng, Z. Wang, and Y.-x. Liu, Topology and
retardation effect of a giant atom in a topological waveguide, Phys. Rev. A
106, 033522 (2022).

\bibitem{CaiPRA104} Q. Y. Cai and W. Z. Jia, Coherent single-photon
scattering spectra for a giant-atom waveguide-QED system beyond the dipole
approximation, Phys. Rev. A 104, 033710 (2021)

\bibitem{GuPRA108} W.j. Gu, H. Huang, Z. Yi, L. Chen, L.h. Sun and H. Tan,
Correlated two-photon scattering in a one-dimensional waveguide coupled to
two- or three-level giant atoms, Phys. Rev. A 108, 053718 (2023).

\bibitem{NoriPRL120} A. F. Kockum, G. Johansson, and F. Nori,
Decoherence-Free Interaction between Giant Atoms in Waveguide Quantum
Electrodynamics, Phys. Rev. Lett. 120, 140404 (2018)

\bibitem{NoriPRL126} X. Wang, T. Liu, A. F. Kockum, H.-R. Li, and F. Nori,
Tunable Chiral Bound States with Giant Atoms,  Phys. Rev. Lett. 126, 043602
(2021).


\bibitem{guoprr2} L.z. Guo, A. F. Kockum, F. Marquardt, and G. Johansson,
Oscillating bound states for a giant atom, Phys. Rev. Res. 2, 043014 (2020).

\bibitem{KianPRA107} Kian Hwee Lim, Wai-Keong Mok, and Leong-Chuan Kwek,
Oscillating bound states in non-Markovian photonic lattices, Phys. Rev. A
107, 023716 (2023).

\bibitem{PRA109} Z. Y. Li and H. Z. Shen, Non-Markovian dynamics with a
giant atom coupled to a semi-infinite photonic waveguide, Phys. Rev. A 109,
023712 (2024).

\end{thebibliography}
\end{document}